\documentstyle[12pt]{article}
\def\be{\begin{equation}}
\def\ee{\end{equation}}
\def\l{\label}

\def\S{{\cal S}}

\def\Z{{}}
\def\H{{\cal H}}
\def\O{{\cal O}}

\def\W{{\cal W}}

 \makeatletter\ifcase\@ptsize\font\teneufm=eufm10
\font\seveneufm=eufm7\font\fiveeufm=eufm5
\font\teneusm=eusm10\font\seveneusm=eusm7
\font\fiveeusm=eusm5\or\font\teneufm=eufm10 scaled
\magstephalf\font\seveneufm=eufm7\font\fiveeufm=eufm5
\font\teneusm=eusm10 scaled\magstephalf
\font\seveneusm=eusm7\font\fiveeusm=eusm5\or \font\teneufm=eufm10
scaled\magstep1\font\seveneufm=eufm7
\font\fiveeufm=eufm5\font\teneusm=eusm10 scaled\magstep1
\font\seveneusm=eusm7\font\fiveeusm=eusm5\fi

\newfam\eufmfam\newfam\eusmfam\textfont\eufmfam=\teneufm
\scriptfont\eufmfam=\seveneufm \scriptscriptfont\eufmfam=\fiveeufm
\textfont\eusmfam=\teneusm\scriptfont\eusmfam=\seveneusm
\scriptscriptfont\eusmfam=\fiveeusm

\def\frak{\ifmmode\let\next\frak@\else
\def\next{\errmessage{Use\string\frak\space only in math
mode}}\fi\next}\def\frak@#1{{\frak@@{#1}}}
\def\frak@@#1{\fam\eufmfam#1}
\def\sh{\ifmmode\let\next\sh@\else
\def\next{\errmessage{Use\string\sh\space only in math
mode}}\fi\next}\def\sh@#1{{\sh@@{#1}}}
\def\sh@@#1{\fam\eusmfam#1}

\ifcase\@ptsize\font\tenmsa=msam10\font\sevenmsa=msam7
\font\fivemsa=msam5\font\tenmsb=msbm10
\font\sevenmsb=msbm7\font\fivemsb=msbm5\or \font\tenmsa=msam10
scaled\magstephalf \font\sevenmsa=msam7\font\fivemsa=msam5
\font\tenmsb=msbm10 scaled\magstephalf
\font\sevenmsb=msbm7\font\fivemsb=msbm5\or \font\tenmsa=msam10
scaled\magstep1\font\sevenmsa=msam7
\font\fivemsa=msam5\font\tenmsb=msbm10 scaled\magstep1
\font\sevenmsb=msbm7\font\fivemsb=msbm5\fi

\newfam\msafam\newfam\msbfam\textfont\msafam=\tenmsa
\scriptfont\msafam=\sevenmsa
\scriptscriptfont\msafam=\fivemsa\textfont\msbfam=\tenmsb
\scriptfont\msbfam=\sevenmsb \scriptscriptfont\msbfam=\fivemsb

\def\Bbb{\ifmmode\let\next\Bbb@\else
\def\next{\errmessage{Use\string\Bbb\space only in math
mode}}\fi\next}\def\Bbb@#1{{\Bbb@@{#1}}}
\def\Bbb@@#1{\fam\msbfam#1}\def\hexnumber@#1{\ifnum#1<10
\number#1\else\ifnum#1=10 A\else\ifnum#1=11 B\else\ifnum#1=12
C\else\ifnum#1=13 D\else\ifnum#1=14 E\else\ifnum#1=15
F\fi\fi\fi\fi\fi\fi\fi}
\def\msa@{\hexnumber@\msafam}\def\msb@{\hexnumber@\msbfam}
\mathchardef\square="0\msa@03

\makeatother

\newcommand{\RR}{{\Bbb R}}
\newcommand{\CC}{{\Bbb C}}

\begin{document}

\begin{titlepage}

\hfill{\tt hep-th/0005274}

\hfill{DFPD00/TH/24}

\begin{center}

{\large\bf Equivalence Postulate and Quantum Origin of
Gravitation}

\vspace{.999cm}

{\large Marco Matone\\} \vspace{.2in}

{\it Department of Physics ``G. Galilei'' -- Istituto Nazionale di
Fisica Nucleare\\ University of Padova, Via Marzolo, 8 -- 35131
Padova, Italy\\ e-mail: matone@pd.infn.it}

\end{center}

\vspace{1.333cm}

\noindent We suggest that quantum mechanics and gravity are
intimately related. In particular, we investigate the quantum
Hamilton--Jacobi equation in the case of two free particles and
show that the quantum potential, which is attractive, may generate
the gravitational potential. The investigation, related to the
formulation of quantum mechanics based on the equivalence
postulate, is based on the analysis of the reduced action. A
consequence of this approach is that the quantum potential is
always non--trivial even in the case of the free particle. It
plays the role of intrinsic energy and may in fact be at the
origin of fundamental interactions. We pursue this idea, by making
a preliminary investigation of whether there exists a set of
solutions for which the quantum potential can be expressed with a
gravitational potential leading term which alone would remain in
the limit $\hbar\rightarrow0$. A number of questions are raised
for further investigation.

\noindent

\vspace{3.333cm}

\noindent PACS: 0.365-w; 0.365.Ca; 0.365.Ta; 04.50.+h


\end{titlepage}\newpage

\section{Introduction}

According to the recently formulated Equivalence Principle (EP),
all physical systems are equivalent under coordinate
transformations 1.--5. It has been shown that the implementation
of such a principle unequivocally leads to the Quantum
Hamilton--Jacobi Equation (QHJE). The latter was first analyzed
independently by Floyd in a series of remarkable papers 6. In 5.
the formulation of Quantum Mechanics (QM) {}from the EP was
extended to higher dimensions and to the relativistic case as
well. This approach suggests that QM and General Relativity (GR)
are two facets of the same medal 1.--5. In this letter we will
argue that QM and GR are intimately related. In particular, we
suggest that gravitation has a purely quantum mechanical origin.

\noindent An outcome of the formulation of QM based on the EP is
that the term $\W\equiv V-E$, with $V$ the potential and $E$ the
energy of the system, corresponds to the inhomogeneous term in the
transformation properties of the state with $\W=\W^0\equiv0$ (see
Refs. 1.--5.). It turns out that this term is of a purely quantum
nature. A related aspect concerns the appearance of fundamental
constants in the QHJE. In particular, the implementation of the EP
leads to the introduction of universal length scales. This has an
important consequence once we take into account that the quantum
potential is always non--trivial. This is a result which follows
{}from a rigorous analysis of the QHJE. Here and throughout this
paper it is important to distinguish between the quantum potential
arising in the approach adopted here and that in the Bohm theory
of QM 7. 8. The two are not in general the same (see Refs. 6. 9.
1.-5.). In particular, it turns out that even in the case of
$\W^0$, the corresponding quantum potential is far {}from being
trivial. This key point is due to the fact that the quantum
reduced action, or quantum Hamiltonian characteristic function, is
always non--trivial. In particular we have \be\S_0\ne
cnst,\l{assolutamentebasilare}\ee which follows as a direct
consequence of the EP 1.--5.

\section{The equivalence postulate}

Before proceeding, let us analyze further the EP. To understand
the basic motivation for its formulation, let us consider, in the
classical framework, two particles of mass $m_A$ and $m_B$ with
relative velocity $v$ 4. For an observer at rest with respect to
the particle $A$, the two systems have reduced actions
$\S^{cl\,A}_0(q_A)=cnst$ and $\S^{cl\, B}_0(q_B)=m_B vq_B$. In
this case, setting $\S^{cl\, A}_0(q_A)= \S^{cl\, B}_0(q_B)$
defines a highly singular coordinate transformation. However, when
the same system is described by an observer in a frame which is
not at rest with respect to either $A$ and $B$, we have that
equating the two reduced actions does not lead to such a
singularity. Thus, this strong singularity disappears if the frame
one uses to describe the systems $A$ and $B$ has a non--zero
velocity with respect to both. For example, if the observer is in
a frame moving with constant acceleration $a$ with respect to the
systems $A$ and $B$, then \be
\tilde\S_0^{cl\,A}(Q_A)={m_A\over3a}(2aQ_A)^{3\over2},\qquad
\tilde\S_0^{cl\,B}(Q_B)={m_B\over3a}(v^2+2aQ_B)^{3\over2},
\l{AB2s}\ee where $Q_A$ ($Q_B$) is the coordinate of particle $A$
($B$) in the accelerated frame. If in describing particle $B$ in
the accelerated frame one uses the coordinate $Q_A$ defined by
$\tilde\S_0^{cl\,A}(Q_A)=\tilde \S_0^{cl\,B}(Q_B)$, then the
resulting dynamics coincides with the one of particle $A$, that is
\be \tilde\S_0^{cl\,B}(Q_B(Q_A))=\tilde\S_0^{cl\,A}(Q_A),
\l{AB3s}\ee which shows that the system $B$, described in terms of
the coordinate $Q_A$, coincides with the system $A$. Hence, in
Classical Mechanics (CM), the equivalence under coordinate
transformations requires choosing a frame in which no particle is
at rest. The existence of a distinguished frame, the one at rest,
seems peculiar as on general grounds what is equivalent under
coordinate transformations in all frames should remain so even in
the one at rest. This leads to postulate that it is always
possible to connect two systems by a coordinate transformation. In
other words, it is natural to require that given two systems with
reduced actions $\S_0$ and $\S_0^v$, there always exists the
``$v$--map'' $q\rightarrow q^v$ defined by 1.--5. \be
\S^v_0(q^v)=\S_0(q). \l{vmapps}\ee The above example shows that
the HJ formalism provides the natural setting to describe physical
systems. The equivalence under coordinate transformations is
somehow in the spirit of general relativity. This property of HJ
theory is still more evident if one notes that the classical HJ
equation itself is obtained by looking for the canonical
transformation of the conjugate variables that leads to the
trivial Hamiltonian $H=0$. Thus, according to CM, since all states
are equivalent, in the sense of the canonical transformations, to
the trivial one, there is a sort of EP. Our view is slightly
different from the one considered in the framework of canonical
transformations. Actually, it is just the above example, in which
equivalence under coordinate transformations always exist except
that in the case one considers the particle at rest, which
suggests the following stronger concept of equivalence. Let us
denote by $\H$ the space of all possible states $\W$. The
equivalence postulate reads 1.--5.

\vspace{.333cm}

\noindent {\it For each pair $\W^a,\W^b\in\H$, there exists a
$v$--transformation such that} \be
\W^a(q)\longrightarrow{\W^a}^v(q^v)=\W^b(q^v). \l{equivalence}\ee

\noindent It has been shown in 1.--5. that the implementation of
the EP unequivocally leads to the quantum HJ equation in any
dimension and in the relativistic case as well.

\section{Fundamental constants and the quantum
potential}

Due to the structure of the QHJE we have that the quantum
potential will in general depend on fundamental constants. Let us
show how these constants arise. We first focus on the
one--dimensional Quantum Stationary Hamilton--Jacobi Equation
(QSHJE). This reads \be {1\over
2m}\left({\partial\S_0(q)\over\partial q}\right)^2+V(q)-E
+{\hbar^2\over4m}\{\S_0,q\}=0, \l{1Q}\ee where
$\{\S_0,q\}=\S_0'''/\S_0'-3(\S_0''/\S_0')^2/2$ denotes the
Schwarzian derivative and $Q={\hbar^2\over4m}\{\S_0,q\}$ is the
quantum potential. The general real solution of (\ref{1Q}) has the
form \be
e^{{2i\over\hbar}\S_0\{\delta\}}=e^{i\alpha}{w+i\bar\ell\over
w-i\ell}, \l{KdT3}\ee where $w=\psi^D/\psi\in\RR$ and
$(\psi^D,\psi)$ are two real linearly independent solutions of the
associated Schr\"odinger equation. Furthermore, we have
$\delta=\{\alpha,\ell\}$, with $\alpha\in\RR$ and
$\ell=\ell_1+i\ell_2$ some integration constants ($\bar\ell$
denoting the complex conjugate of $\ell$). Observe that $\ell_1\ne
0$, which is equivalent to having $\S_0\ne cnst$, is a necessary
condition to define the term $\{\S_0,q\}$ in the QSHJE.

\noindent There is a simple reason why fundamental constants
should be hidden in $\ell$. To see this, consider the
Schr\"odinger equation in the trivial case $\W^0(q^0)\equiv0$,
that is $\partial_{q^0}^2\psi=0$. Two linearly independent
solutions are $\psi^D=q^0$ and $\psi=1$. Now a basic aspect of the
formulation is manifest duality between real pairs of linearly
independent solutions 1.--5. This is a fact which is strictly
related to the Legendre duality first observed in 10. and further
investigated in 11.--14. Thus, whereas in the standard approach
one usually considers only one solution of the Schr\"odinger
equation, {\it i.e.} the wave--function itself, in the present
formulation both $\psi^D$ and $\psi$ enter the relevant formulas.
This leads to expressions containing linear combinations of
$\psi^D$ and $\psi$, typically $\psi^D+i\ell \psi$ that for
$\psi^D=q^0$ and $\psi=1$ reads $q^0+i\ell_0$, so
$\ell_0\equiv\ell$ should have the dimensions of a length. The
fact that $\ell$ has the dimensions of a length is true for any
state. This follows from the observation that the ratio
$w=\psi^D/\psi$ is a M\"obius transformation of the trivializing
map transforming any state to $\W^0$ 1.--5. Hence $w$, and
therefore $\ell$, has the dimensions of a length.

\noindent Since $\ell_0$ enters the QSHJE with $\W^0\equiv0$, the
system does not provide any dimensionful quantity. This implies
that we have to introduce some fundamental lengths. Let us show
this in some detail. The reduced action $\S_0^0$ corresponding to
the state $\W^0$ is \be
e^{{2i\over\hbar}\S_0^0\{\delta\}}=e^{i\alpha}{q^0+i\bar\ell_0\over
q^0-i\ell_0}, \l{KdT3bisse}\ee and the conjugate momentum
$p_0=\partial_{q^0}\S^0_0$ has the form \be p_0=\pm {\hbar
(\ell_0+\bar\ell_0)\over 2|q^0-i\ell_0|^2}. \l{powP}\ee A property
of $p_0$ is that it vanishes only for $q^0\rightarrow\pm\infty$.
Furthermore, $|p_0|$ reaches its maximum at $q^0=-{\rm Im}\,
\ell_0$ \be |p_0(-{\rm Im}\, \ell_0)|={\hbar\over {\rm Re}\,
\ell_0}. \l{pdoiw}\ee Since ${\rm Re}\, \ell_0\ne0$, $p_0$ is
always finite. Thus, ${\rm Re}\, \ell_0\ne0$ provides a sort of
ultraviolet cutoff. This is a property which extends to arbitrary
states. Actually, the conjugate momentum has the form \be p={\hbar
W (\ell+\bar\ell)\over2\left|\psi^D-i\ell\psi\right|^2},
\l{momentino}\ee where $W=\psi'\psi^D-\psi^{D'}\psi$ is the
Wronskian. Since $W$ is a non--vanishing constant, it follows that
$\psi^D$ and $\psi$ cannot have common zeroes, and by ${\rm Re}\,
\ell\ne0$ we see that $p$ is finite $\forall q\in \RR$. Therefore,
the EP implies an ultraviolet cutoff on the conjugate momentum.

\noindent In Refs. 2. and 4. it has been shown that fundamental
constants also arise in considering the classical limit. In
particular, one first considers \be \lim_{\hbar\rightarrow0}p_0=0,
\l{prova23}\ee and note that ${\rm Im}\,\ell_0$ in (\ref{powP})
can be absorbed by a shift of $q^0$. Hence, in (\ref{prova23}) we
can set ${\rm Im}\,\ell_0=0$ and distinguish the cases $q^0\ne0$
and $q^0=0$. From (\ref{prova23}) \be p_0
{}_{\stackrel{\sim}{\hbar\rightarrow 0}}
\left\{\begin{array}{ll} \hbar^{\gamma+1}, & q_0\ne 0,\\
\hbar^{1-\gamma}, & q_0=0, \end{array}\right. \l{Vu1fy9}\ee where
$-1<\gamma<1$ with $\gamma$ defined by ${\rm Re}\,\ell_0
{}_{\stackrel{\sim}{\hbar\rightarrow 0}}\hbar^{\gamma}$. The are
not many fundamental lengths in nature. In particular, we note
that a fundamental length satisfying this condition on the power
of $\hbar$ is the Planck length $\lambda_p=\sqrt{\hbar G/c^3}$,
while the Compton length is excluded by the condition $\gamma<1$.
Also, as we will see in considering the $E\rightarrow0$ and
$\hbar\rightarrow0$ limits for the free particle of energy $E$,
the natural choice is just the Planck length. With this choice of
${\rm Re}\, \ell_0$ the maximum of $|p_0|$ is \be |p_0(-{\rm Im}\,
\ell_0)|=\sqrt{c^3\hbar\over G}. \l{pdoiwtrisse}\ee Setting ${\rm
Im}\, \ell_0=0$ and ${\rm Re}\, \ell_0=\lambda_p$, the quantum
potential associated to the trivial state $\W^0$ is \be
Q^0={\hbar\over 4m}\{\S_0^0,q^0\}=-{\hbar^3 G
\over2mc^3}{1\over|q^0-i\lambda_p|^4}. \l{oix9ui87}\ee There are
two basic aspects in this expression. Firstly, the gravitational
constant $G$ results from ensuring consistency with the classical
limit. We saw that this arises naturally as a consistency
condition. Furthermore, $Q^0$ is negative definite. Thus, even if
we are still considering a one--dimensional system, we are
starting to see some motivation for the emergence of the
gravitational interaction. In particular, note that this analysis
is essentially the same of the one in three dimensional space as
in the case of a free particle we can consider a reduced action of
the form $\S_0(x)+\S_0(y)+\S_0(z)$, which can always be chosen as
a possible solution of the QHSJE when the potential has the form
$V(x,y,z)=V_1(x)+V_2(y)+V_3(z)$. We also note that the fact that
we are in the framework of non--relativistic QM, does not exclude
the appearance of $c$ in the integration constants of the QSHJE
(an example of the appearance of $c$ in QM is the
non--relativistic treatment of an electron in a magnetic field).

\noindent The appearance of fundamental constants can also be seen
by considering the $\hbar\rightarrow0$ and $E\rightarrow0$ limits
2. for the conjugate momentum of a free particle of energy $E$ \be
p_E=\pm{\hbar(\ell_E+\bar\ell_E)\over2\left|k^{-1}\sin(kq)-i\ell_E
\cos(kq)\right|^2}, \l{CS15}\ee where $k=\sqrt{2mE}/\hbar$ and
$\ell_E$ is the integration constant of the QSHJE. We should
require that (see Refs. 2. and 4.) \be
\lim_{\hbar\rightarrow0}p_E=\pm\sqrt{2mE}, \l{bos1S11b}\ee and \be
\lim_{E\rightarrow0}p_E=p_0=\pm{\hbar(\ell_0+\bar\ell_0)
\over2|q-i\ell_0|^2}. \l{bisCS11b}\ee However, we see that the
term $\ell_E\cos(kq)$ in Eq.(\ref{CS15}) is ill--defined in the
$\hbar\rightarrow0$ limit, a problem which has been recently
considered also by Floyd 15. Thus, the existence of the classical
limit implies some condition on $\ell_E$. In particular, in order
to reach the classical value $\sqrt{2mE}$ in the
$\hbar\rightarrow0$ limit, the quantity $\ell_E$ should depend on
$E$. In Refs. 2. and 4. it has been shown that \be
\ell_E=k^{-1}e^{-\alpha(x_p^{-1})}+e^{-\beta(x_p)}\ell_0,
\l{prova10}\ee where $x_p=k\lambda_p=\sqrt{2mEG/\hbar c^3}$ and
\be \alpha(x_p^{-1})=\sum_{k\geq1}\alpha_kx_p^{-k},\qquad
\beta(x_p)=\sum_{k\geq1}\beta_k x_p^k. \l{alphapp}\ee It follows
that \be p_E=\pm{2k^{-1}\hbar e^{-\alpha(x_p^{-1})}+\hbar
e^{-\beta(x_p)}(\ell_0+\bar\ell_0)\over2\left|k^{-1}\sin(kq)-i
\left(k^{-1}e^{-\alpha(x_p^{-1})}+e^{-\beta
(x_p)}\ell_0\right)\cos(kq)\right|^2}. \l{prova21bbvxx}\ee The
function $\alpha(x_p^{-1})$ is constrained by the conditions \be
\lim_{\hbar\rightarrow0} e^{-\alpha(x_p^{-1})}=1,\qquad
\lim_{E\rightarrow0}E^{-1/2} e^{-\alpha(x_p^{-1})}=0,
\l{prova6}\ee whereas for $\beta(x_p)$ we have \be
\lim_{\hbar\rightarrow0} \hbar^{-1}e^{-\beta(x_p)}\ell_0=0.
\l{prova21}\ee One of the conditions we used to derive the above
formulas concerns the existence of the classical limit. In this
context we should observe that it may be that the classical
expressions themselves may contain further terms which do not
depend on $\hbar$. As an example, observe that two free particles
should always contain the gravitational potential as intrinsic
interaction. This seems to be connected with the residual
indeterminacy discussed in 15. The aim of the present paper is to
investigate the possibility that such interaction may be a
consequence of the quantum potential.

\noindent The appearance of the Planck scale in the hidden
constants has been considered in Refs. 2. and 4. This seems
related to the 't Hooft's approach 16. Possible connections have
been considered by Floyd 15. and in 5.

\section{The cocycle condition and the quantum
nature of interactions}

Let us further consider the nature of the EP itself. We introduce
the notation \be J_{ki}={\partial q_i\over\partial q^v_k},
\l{zummoloa12}\ee and \be (p^v|p)={\sum_k
p_k^{v^2}\over\sum_kp_k^2}={p^tJ^tJp\over p^tp}. \l{natura2}\ee
The only possibility to reach any other state $\W^v\ne0$ starting
{}from $\W^0$ is that it transforms with an inhomogeneous term
1.--5. \be \W^v(q^v)=(p^v|p^a)\W^a(q^a)+\Z(q^a;q^v),
\l{azzoyyyxxaa10bbbb}\ee and \be
Q^v(q^v)=(p^v|p^a)Q^a(q^a)-\Z(q^a;q^v), \l{azzo2yyyxxaa10bbbb}\ee
where $\Z(q^a;q^v)$ denotes a still undefined function which
depends on $q^a$ and $q^v$. Let us denote by $a,b,c,\ldots$ a set
of different $v$--transformations. Comparing
\be\W^b(q^b)=(p^b|p^a)\W^a(q^a)+\Z(q^a;q^b)=\Z(q^0;q^b),
\l{ganzate}\ee with the same formula with $q^a$ and $q^b$
interchanged we have \be \Z(q^b;q^a)=-(p^a|p^b)\Z(q^a;q^b).
\l{inparticolare}\ee More generally, comparing \be
\W^b(q^b)=(p^b|p^c)\W^c(q^c)+\Z(q^c;q^b)=(p^b|p^a)\W^a(q^a)+
(p^b|p^c)\Z(q^a;q^c)+\Z(q^c;q^b), \l{zummoloa14}\ee with
(\ref{ganzate}) we obtain the basic cocycle condition \be
\Z(q^a;q^c)=(p^c|p^b)\left[\Z(q^a;q^b)-\Z(q^c;q^b)\right],
\l{cociclo3}\ee which is the essence of the EP. In particular,
this condition unequivocally leads to determine the correction to
the CHJE. In doing this, one shows that Eq.(\ref{cociclo3})
implies a basic M\"obius invariance of $\Z(q^a;q^b)$. The
$\W^0\equiv0$ state plays a special role. Setting $\W^a=\W^0$ in
Eq.(\ref{azzoyyyxxaa10bbbb}) yields $\W^v(q^v)=\Z(q^0;q^v)$. Thus,
in general \be \W(q)=\Z(q^0;q), \l{ddazzoyyyxxaa10bbbb}\ee so
that, according to the EP (\ref{equivalence}), all states
correspond to the inhomogeneous part in the transformation of the
$\W^0$ state induced by some $v$--map. Since the inhomogeneous
part has a purely quantum origin, we conclude that the Equivalence
Postulate implies that interactions have a purely quantum origin.

\noindent The role of the quantum potential as responsible for
interactions can be made clearer from the observation that the EP
implies \be \W^v(q^v)+Q^v(q^v)=(p^v|p)\left(\W(q)+Q(q)\right).
\l{yyyxxaa10bbbb}\ee Then, taking $\W=\W^0\equiv0$ and omitting
the superscript $v$, we have \be \W(q)=(p|p^0)Q^0(q^0)-Q(q),
\l{originequantistica}\ee showing that any potentials can be
expressed in quantum terms.

\noindent In 5. it has been observed that there is a hidden
antisymmetric tensor in QM which arises from the continuity
equation. We also note that in the one--dimensional case, the
freedom deriving from the underlying hidden tensor one meets in
the higher dimensional case reflects itself in the appearance of
the integration constants. These are related to the $SL(2,\CC)$
symmetry \be e^{2i\S_0/\hbar}\longrightarrow
{Ae^{2i\S_0/\hbar}+B\over Ce^{2i\S_0/\hbar}+D},\l{ofinw}\ee of the
equation \be \{e^{2i\S_0/\hbar},q\}=-4m\W/\hbar^2, \l{odiP}\ee
which is equivalent to the QSHJE (\ref{1Q}). In particular, as we
said, there is a complex integration constant $\ell$ which is
missing in the Schr\"odinger equation. Changing this constant
corresponds to a M\"obius transformation (\ref{ofinw}). While this
leaves $\W$ unchanged, it mixes the quantum potential and the
kinetic term. Thus the quantum potential is essentially
parameterized by $SL(2,\CC)$ transformations in which the
constants $A,B,C$ and $D$ depend, by dimensional analysis and
consistency of relevant limits considered above, on fundamental
constants. We may expect that these constants and the above
M\"obius transformations in three and four dimension (for the
relativistic generalization) should be related to fundamental
interactions.\footnote{It is worth mentioning that the geometrical
building block of string theory, which also explains why string
theory includes gravity, is the thrice punctured Riemann sphere.
The latter can be characterized just by the basic $SL(2,\CC)$
M\"obius symmetry related to the arbitrariness of the position of
the punctures.}

\section{The two--particle model}

The above investigation suggests considering that the quantum
potential $Q$ is at the origin of the interactions. Thus, it may
be that the constants defining $Q$ depend on the intrinsic
properties of the particles. This would lead to different possible
forms of $Q$ and therefore of the admissible interactions. As we
saw, there are subtle questions concerning the classical limit.
Similarly, one should consider the relativistic case as it may
lead to results which may remain hidden if considered directly in
the non--relativistic case. Similarly, at least in the case of
gravitational interaction, one should consider the analysis of
macroscopic objects to take into account possible collective
effects. Nevertheless, the above suggestions indicate that it is
worth studying the case of two free particles and then looking at
the possible structure of the quantum potential.

\noindent In the case of two free particles of energy $E$ and
masses $m_1$ and $m_2$, the QSHJE reads \be {1\over
2m_1}(\nabla_1\S_0)^2+{1\over
2m_2}(\nabla_2\S_0)^2-E-{\hbar^2\over 2m_1}{\Delta_1 R\over
R}-{\hbar^2\over 2m_2}{\Delta_2 R\over R}=0. \l{QSHJEHDcondue}\ee
The continuity equation is \be {1\over
m_1}\nabla_1\cdot(R^2\nabla_1\S_0)+ {1\over
m_2}\nabla_2\cdot(R^2\nabla_2\S_0)=0. \l{conteqHDcondue}\ee Next,
we set \be r=r_1-r_2,\qquad r_{c.m.}={m_1r_1+m_2r_2\over
m_1+m_2},\qquad m={m_1m_2 \over m_1+m_2}, \l{pcwam}\ee where $r_1$
and $r_2$ are the ray vectors of the two particles. With respect
to the new variables the equations (\ref{QSHJEHDcondue}) and
(\ref{conteqHDcondue}) have the form \be {1\over
2(m_1+m_2)}(\nabla_{r_{c.m.}}\S_0)^2+{1\over 2m}
(\nabla\S_0)^2-E-{\hbar^2\over 2(m_1+m_2)}{\Delta_{r_{c.m.}}
R\over R}-{\hbar^2\over 2m}{\Delta R\over R}=0,
\l{QSHJEHDcondueetre}\ee \be {1\over m_1+m_2}\nabla_{
r_{c.m.}}\cdot(R^2\nabla_{r_{c.m.}}\S_0)+{1\over
m}\nabla\cdot(R^2\nabla\S_0)=0, \l{conteqHDcondueetre}\ee where
$\nabla$ ($\nabla_{r_{c.m.}}$) and $\Delta$ ($\Delta_{r_{c.m.}}$)
are the gradient and Laplacian with respect to the components of
the vector $r$ ($r_{c.m.}$). These equations can be decomposed
into the equations for the center of mass $r_{c.m.}$ and those for
the relative motion. We will concentrate on the latter. It
satisfies the QSHJE \be {1\over 2m}(\nabla\S_0)^2-E-{\hbar^2\over
2m}{\Delta R\over R}=0, \l{QSHJEHD}\ee and the continuity equation
\be \nabla \cdot(R^2\nabla\S_0)=0. \l{conteqHD}\ee In 5 it has
been stressed that the continuity equation implies \be
R^2\partial_i\S_0=\epsilon_i^{\;\,i_2\ldots i_D}\partial_{i_2}
F_{i_3\ldots i_D}, \l{errequadro}\ee where $F$ is a $(D-2)$--form.
In the 3D case $R^2 \partial_i\S_0$ is the curl of a vector that
we denote by $B$ \be \nabla\S_0=R^{-2}\nabla\times B.
\l{JimiHendrix}\ee The QSHJE (\ref{QSHJEHD}) reduces to the
``canonical form'' \be j^2={\hbar^2}R^3\Delta R+2mER^4,
\l{aopdfc}\ee where $j^2\equiv j_kj^k$ with \be j=\nabla\times B.
\l{FrankZappa}\ee Using the identity $(a\times b)\cdot (c\times
d)= (a\cdot c)(b\cdot d)- (a\cdot d)(b\cdot c)$, Eq.(\ref{aopdfc})
reads \be \Delta B^2 -(\nabla B)^2={\hbar^2}R^3\Delta R+2mER^4.
\l{aopdfcduetto}\ee It is worth stressing that $j$ resembles the
usual current. However, besides the mass term, we stress again
that here $R$ and $\S_0$ are not in general the ones one obtains
identifying $Re^{{i\over\hbar}\S_0}$ with the wave--function.
Nevertheless, by construction we have that
$\psi=Re^{{i\over\hbar}\S_0}$ solves the Schr\"odinger equation.
Thus, we have \be j={i\hbar\over
2}(\psi\nabla\overline\psi-\overline\psi\nabla\psi),
\l{correntizia}\ee that, in the case in which $\psi$ is the
wave--function, coincides, upon dividing by $m$, with the usual
quantum mechanical current.

\noindent We have seen that two free particles have a non--trivial
quantum potential whose structure depends on the field $B$. In the
following we will use the above results in order to investigate
whether this potential may in fact have a gravitational leading
behavior.

\section{Gravitational interaction and quantum
potential}

After summarizing the main results so far, we will write down the
differential equation that the quantum potential should satisfy in
order to obtain the gravitational interaction. We will then
investigate in some detail such an equation.

\noindent The aim of the previous sections was to show the main
aspects suggesting that the quantum potential is at the origin of
fundamental interactions. Even if these aspects have been
discussed in detail, it is useful to collect them together before
formulating the hypothesis and then deriving the relevant
equations.

\begin{enumerate}

\item The EP implies that the reduced action is always non--trivial.
In particular, this is true also for the free particle of
vanishing energy. Furthermore, if $\psi\propto\overline\psi$, such
as in the case of the wave--function for bound states, then
$\psi=R\left(A
e^{-{i\over\hbar}\S_0}+Be^{{i\over\hbar}\S_0}\right)$, with
$\psi\propto\overline\psi$ giving $|A|=|B|$. Thus there is no
track of the condition $\S_0=cnst$. On the other hand, this cannot
be a solution of the QSHJE and would give an inconsistent
classical limit. Remarkably, this answers the objections
concerning the classical limit posed by Einstein. He just noticed
that for a particle in a box the identification of the
wave--function with $Re^{{i\over\hbar}\S_0}$ gives $\S_0=cnst$ and
this cannot reproduce, in the $\hbar\rightarrow0$ limit, the
non--trivial $\S_0^{cl}$. This result has been previously derived
by Floyd in a series of important papers 6. Related aspects have
also been considered in the interesting papers by
Reinisch\footnote{I am grateful to G. Reinisch who informed me
that the argument about the unphysical $\hbar\to0$ limit was
explicitly used by Einstein (see pg.243 of Holland's book 8.).} 9.

\item This property of the reduced action implies the existence of
an intrinsic potential energy which, like the rest mass of special
relativity, is universal. In particular, the quantum potential is
always non--trivial. This is different from the standard approach
where there are examples in which $Q=0$ so that the QHJE would
coincide with the classical one.

\item The existence of the classical limit implies that the
quantum potential depends, through the hidden initial conditions
coming from the QSHJE, on fundamental length scales which in turn
depend on $\hbar$. It is a basic fact that these initial
conditions are missing in the Schr\"odinger equation. In
particular, the emergence of the Planck length, and therefore of
Newton's constant, arises from considering the classical limit for
the free particle of vanishing energy.

\item It can be seen in the formulation that the quantum
potential provides particle's response to an external
perturbation. For example, in the case of tunnelling, the
attractive nature of the quantum potential guarantees the reality
of the conjugate momentum and therefore of the velocity field
$v=1/\partial_Ep\ne p/m$ (see Refs. 6. and 4.). More precisely,
inside the barrier the quantum potential decreases its value in
such a way that $(\partial_q\S_0)^2$ remains positive definite. As
a consequence, the role of this internal energy, which is a
property of all forms of matter, should manifest itself through
effective interactions depending on the above fundamental
constants.

\item The fundamental implication of the EP is the cocycle
condition (\ref{cociclo3}). In particular, from this condition,
one obtains an expression for the interaction terms which is
purely of quantum origin.

\item The fact that QM arises from an EP which is reminiscent of
Einstein's EP strongly indicates a deep relation between
gravitation and QM itself.

\end{enumerate}

\noindent The most characteristic property of the quantum
potential is its universal nature: it is a property possessed by
all forms of matter. On the other hand, we know that such a
property is the one characterizing gravity. Therefore, if we write
down the classical equations of motion for a pair of particles, we
should always include, already at the classical level, the
gravitational interaction. Furthermore, the quantum potential for
a free particle is negative definite. This should be compared with
the attractive nature of gravity.

\section{The quantum potential with the
gravitational potential as a leading term}

The above remarks suggest formulating the hypothesis that the
quantum potential is in fact at the origin of gravitation. Thus we
look for solutions of the QSHJE leading to the classical HJ
equation for the gravitational interaction. In particular, we
should investigate whether in the case of two free particles the
quantum potential \be Q=-{\hbar^2\over2m}{\Delta R\over R},
\l{erquantupotentia}\ee admits the form \be Q=V_{G},
\l{quelliderprlnannocapitoncazzo}\ee with $V_G$ reducing to the
Newton potential in the $\hbar\rightarrow0$ limit \be
\lim_{\hbar\rightarrow0}V_G=-G{m_1m_2\over r}. \l{erlimite}\ee If
such a solution exists then, in the limit $\hbar\rightarrow0$,
Eq.(\ref{QSHJEHD}) corresponds to the HJ equation for the
gravitational potential \be {1\over
2m}(\nabla\S_0^{cl})^2-G{m_1m_2\over r}-E=0. \l{QSHJEHDbisse}\ee
Summarizing, the above problem corresponds to finding all the
possible $R$ satisfying the equation \be {\hbar^2\over2m}{\Delta
R\over R}=-V_G=G{m_1m_2\over r}+{\O}(\hbar),
\l{Shutupandplayyerguitar}\ee where the higher order terms
${\O}(\hbar)$ will generally depend on $r$, such that $R$ and
$\S_0$ satisfy Eqs.(\ref{QSHJEHD}) and (\ref{conteqHD}). Let us
consider the set ${\cal R}=\{R|sol. \; of\;
(\ref{Shutupandplayyerguitar})\}$. The above problem is equivalent
to find the set ${\cal B}=\{B|sol. \; of\; (\ref{aopdfcduetto})\;
with\; R\in{\cal R}\}$ (recall that if $R$ and $B$ solve
Eq.(\ref{aopdfcduetto}), then $\nabla\S_0=R^{-2}\nabla\times B$ is
solution of the QSHJE and of the continuity equation). It follows
that the set of possible potentials with gravitational behavior
$r^{-1}$ is given by \be {\cal V}_G=\left\{-{\hbar^2\over
2m}{\Delta R\over R}\,\bigg|R\in{\cal R}_G\right\}, \l{VGG}\ee
where ${\cal R}_G=\{R|R\in{\cal R}, B\; exists\}$. In other words,
we have to find all the possible $R$ satisfying
(\ref{Shutupandplayyerguitar}) and then restricting to those for
which there exists a field $B$ satisfying (\ref{aopdfcduetto}).
This would fix the set of admissible potentials ${\cal V}_G$ to be
investigated. Note that the fact that the higher order terms in
(\ref{Shutupandplayyerguitar}) are not fixed implies that ${\cal
R}$ has infinitely many elements. This set identifies infinitely
many equations of the kind (\ref{aopdfcduetto}), one for each
$R\in {\cal R}$. Thus, on general grounds, one should expect that
the set ${\cal B}$, and therefore ${\cal V}_G$, be non--trivial.

\section{The spherical case}

While an adequate treatment of the above problem will be
considered in a future publication, here we consider some related
preliminary aspects. By introducing the $B$ field we saw that it
should be possible to a find a solution to the two--particle
model. However, a more effective way of considering such a problem
seems to reformulate it as follows. First we note that by
(\ref{QSHJEHD}) and (\ref{Shutupandplayyerguitar}) we have that
$\S_0$ should satisfy the equation \be {1\over
2m}(\nabla\S_0)^2=E+G{m_1m_2\over r}+{\O}(\hbar). \l{QSHJEHD4}\ee
Thus, instead of finding first the possible $R\in{\cal R}$, it
seems convenient to solve Eq.(\ref{QSHJEHD4}) which looks simpler
than Eq.(\ref{Shutupandplayyerguitar}). A general solution of this
equation would involve terms depending also on $\theta$ and
$\phi$. However, the simplest situation is when $\S_0$ is a
function of $r$. In this case $\nabla\S_0=\hat
r\partial_r\S_0(r)$, where $\hat r$ is the unit vector along $r$.
Eq.(\ref{QSHJEHD4}) becomes \be {1\over
2m}(\partial_r\S_0)^2=E+G{m_1m_2\over r}+{\O}(\hbar),
\l{QSHJEHD5}\ee and the continuity equation reads
$\nabla\cdot(R^2\hat r \partial_r\S_0)=0$, giving \be R={1\over
r\sqrt{\partial_r\S_0}}.\l{parac}\ee Since the radial part of the
Laplacian is $r^{-1}\partial^2_rr$, we have that the QSHJE
(\ref{QSHJEHD}) becomes \be {1\over
2m}(\partial_r\S_0)^2-E+{\hbar^2\over4m}\{\S_0,r\}=0.
\l{basilarissima}\ee Formally this equation is the
one--dimensional QSHJE for a free particle on the non--negative
part of the real axis. Therefore, by Eq.(\ref{CS15}) we have \be
\partial_r\S_0=\pm{\hbar(\ell_E+\bar\ell_E)\over2\left|k^{-1}
\sin(kr)-i\ell_E\cos(kr)\right|^2}. \l{CS15222}\ee To establish
the right asymptotic we should handle the indeterminacy discussed
above and eliminated by a suitable choice of the constant
$\ell_E$. In particular, while in the previous case the structure
of $\ell_E$ was fixed by requiring that
$p_E\rightarrow\pm\sqrt{2mE}$ as $\hbar\rightarrow0$, we should
now investigate the full functional structure of the right hand
side of (\ref{CS15222}) at the different scales defined by the
parameters $\ell_E$, $\hbar$, $m$ and $E$.

\noindent We now consider the general case by adding to $\S_0$ the
dependence on $\theta$ and $\phi$ and then studying the possible
appearance of the $r^{-1}$ term. Setting
$\psi=Re^{{i\over\hbar}\S_0}$, which is a solution of the
Schr\"odinger equation, we have $\S_0={\hbar\over
2i}\ln(\psi/\overline\psi)$, so that \be
(\nabla\S_0)^2=-{\hbar^2\over 4|\psi|^4}\sum_{j=1}^3(\overline
\psi\partial_j\psi -\psi\partial_j\overline \psi)^2,\l{iug}\ee
where $\partial_1=\partial_x$, $\partial_2=\partial_y$ and
$\partial_3=\partial_z$. Since $\psi$ solves the free
Schr\"odinger equation, we have \be
\psi=\sum_{l=0}^\infty\sum_{m=-l}^l\sum_{j=1}^2 c_{lmj}
R_{klj}(r)Y_{lm}(\theta,\phi),\l{p43eik}\ee where the
$Y_{lm}(\theta,\phi)$ denote the spherical harmonics and \be
R_{kl1}= (-1)^l2{r^l\over k^l} \left({1\over
r}\partial_r\right)^l{\sin kr\over r}, \qquad R_{kl2}=
(-1)^l2{r^l\over k^l} \left({1\over r}\partial_r\right)^l{\cos
kr\over r}.\l{oiweudxch}\ee These are linearly independent
solutions of the radial part of the Schr\"odinger equation \be
R_{klj}''+ {2\over r} R_{klj}'+\left[k^2-{l(l+1)\over
r^2}\right]R_{klj}=0.\l{oiuuh}\ee We are studying this equation at
$r>0$. In this respect note that the singularity of $R_{kl2}$ at
$r=0$ would give a term $\delta(r)$ in the right hand side of
(\ref{oiuuh}). In spherical coordinates we have
\be(\nabla\S_0)^2=-{\hbar^2\over 4|\psi|^4}\left[(\overline
\psi\partial_r\psi -\psi\partial_r\overline\psi)^2+{1\over r^2}
(\overline\psi\partial_\theta\psi -\psi\partial_\theta\overline
\psi)^2+{1\over r^2\sin^2\theta} (\overline\psi\partial_\phi\psi
-\psi\partial_\phi\overline\psi)^2\right].\l{iug2}\ee The
properties of this expression will be considered elsewhere.
However, as a preliminary step, we consider the first term in the
square bracket in (\ref{iug2}). Note that
\be\nabla=\left(\partial_r,{1\over
2r}(e^{-i\phi}l_+-e^{i\phi}l_-), {i\over
r\sin\theta}l_z\right),\l{oixjqa}\ee where $l_{\pm}=l_x\pm i
l_y=e^{\pm i\phi}(\partial_\theta+ i\cot \theta\partial_\phi)$,
with $l_x$, $l_y$ and $l_z$ denoting the components of the angular
momentum operator. Since $l_+Y_{lm}=a_{lm}Y_{lm+1}$ and
$l_-Y_{lm}=a_{lm-1}Y_{lm-1}$, with
$a_{lm}\equiv\sqrt{(l+m+1)(l-m)}$, we have \be \nabla\psi=
\sum_{\{lmj\}}\left(c_{lmj}R_{klj}'Y_{lm}, {1\over
2r}c_{lmj}R_{klj}(e^{-i\phi}a_{lm}Y_{lm+1}-e^{i\phi}a_{lm-1}Y_{lm-1}),
{i\over r\sin\theta}c_{lmj}R_{klj}mY_{lm}\right),\l{oieufgh}\ee
where
$\sum_{\{lmj\}}\equiv\sum_{l=0}^\infty\sum_{m=-l}^l\sum_{j=1}^2$.
Finally, using $R_{klj}'=lr^{-1}R_{klj}-kR_{kl+1j}$, we have \be
(\overline\psi\partial_r\psi
-\psi\partial_r\overline\psi)^2=-4{\Im}^2\left(\sum_{\{lmj\}}
\sum_{\{l'm'j'\}}\overline c_{l'm'j'}R_{kl'j'}\overline
Y_{l'm'}c_{lmj}(lr^{-1}R_{klj}-kR_{kl+1j})Y_{lm}
\right).\l{pqwoDFJ}\ee

\section{Conclusions}

Let us conclude by observing that the aim of the present
investigation is to propose a possible quantum origin of the
gravitational interaction. In particular, we made a preliminary
investigation of the problem of finding the set ${\cal V}_G$ of
potentials with gravitational leading term originated by the
quantum potential of two free particles. The general solution
seems to be of mathematical interest and will be considered in a
future publication. In this context we would like to mention the
Schr\"odinger--Newton equation 17. which concerns a problem
reminiscent of the one introduced in this paper. Finally, we would
like to mention that geometrical aspects related to the quantum HJ
equation have been considered also in 18. and references therein.

\vspace{.333cm}

\noindent {\bf Acknowledgements}. It is a pleasure to thank the
anonymous Referees for interesting comments and D. Bellisai, G.
Bertoldi, R. Carroll, A.E. Faraggi, E.R. Floyd, J.M. Isidro, P.A.
Marchetti, M. Mariotti, L. Mazzucato, P. Pasti, G. Reinisch, P. Sergio
and M. Tonin, for stimulating discussions. Work supported in part by the
European Commission TMR programme ERBFMRX--CT96--0045.

\end{document}